\documentclass[%
 aip,
 pra,
 amsmath,amssymb,
 preprint,%
]{revtex4-2}

\usepackage{amssymb,amsmath,amsfonts}
\usepackage{graphicx,graphics}
\usepackage{epsfig}
\usepackage{blkarray}
\usepackage{epstopdf}
\usepackage[FIGBOTCAP]{subfigure}
\usepackage{amsfonts}
\usepackage{bm,bbm}
\usepackage{amssymb,amsmath,amsfonts}
\usepackage{mathrsfs}
\usepackage{calc}
\usepackage{epsfig}
\usepackage{float}
\usepackage{color}
\usepackage{epstopdf}
\usepackage{bm,amssymb}
\usepackage{graphicx,color}
\usepackage{comment}
\usepackage{soul}
\usepackage{amsmath}
\newcommand{\be}{\begin{equation}}
\newcommand{\ee}{\end{equation}}

\newcommand{\beqar}{\begin{eqnarray}}
\newcommand{\eeqar}{\end{eqnarray}}
\newcommand{\bcen}{\begin{center}}
\newcommand{\ecen}{\end{center}}

\DeclareUnicodeCharacter{2212}{-}

\begin{document}

\title{{\color{black}  Conditions for enhancement of gas phase chemical reactions inside a dark microwave cavity}}
\author{Nimrod Moiseyev$^{1,2,3}$ \footnote{\tt nimrod@technion.ac.il https://nhqm.net.technion.ac.il/} }

\affiliation{Schulich Faculty of Chemistry$^1$, Faculty of Physics$^2$ and Solid State Institute$^3$, Technion-Israel Institute of Technology, Haifa 32000, Israel}

\begin{abstract}
Enhancing chemical reactions, such as  $A+B \to [\textit{activated complex}]^\# \to C+D$, in gas phase through its coupling to quantized electromagnetic modes in a dark cavity is investigated.
The main result is that the enhancement of the reaction rate by a dark cavity occurs for asymmetric reactions (products different from reactants).  {\color{black} In addition to the cavity being dark, also the reactants are required to be in their ground electronic and vibrational states, i.e., the whole system should be prepared as dark.}
{Theoretical derivation, utilizing the non-Hermitian formalism of quantum mechanics (NHQM),
provides conditions and guidelines for selecting the proper type of reactions that can be enhanced by a dark cavity.
Nevertheless, the time-dependent simulations of such experiments can be carried out using the standard (Hermitian) scattering theory (while utilizing the conditions derived via NHQM).}
We believe that this work opens a gate to new types of studies and hopefully helps to close the gap between theory and experiments in this fascinating, relatively new field of research.
As an example, we demonstrate that the asymmetric reaction rates of $O+D_2\to [ODD]^{\#} \to OD+D$ and $H+ArCl \to [ArHCl]^{\#} \to H+Ar+Cl$ can be enhanced by a dark cavity.
On the other hand, an effect of the dark cavity on the symmetric reaction of hydrogen exchange in methane is predicted to be negligible. {\color{black} The applications of the theory presented here are applied to chemical reactions inside a dark microwave cavity. However, the theory is applicable to any other type of cavity and is not limited to microwave cavities only.}

\end{abstract}

\maketitle
\section{Motivation}

In 2012, Thomas  Ebbesen  and his coworkers\cite{Ebbesen2012} showed for the first time that the vacuum field can suppress  the rate of a chemical reaction. This work open the field of polaritonic chemistry.
In 2019 Ebbesen and Jino George teams reported on increase in the reaction rate by an order of magnitude due to  the catalytic effect of vibrational strong coupling on the solvolysis of para-nitrophenyl acetate\cite{Enhance-DARK-CAVITY-2019}, and in 2022 the  enhancement of the rate of an esterification reaction has been reported\cite{Enhance-DARK-CAVITY-2022}.
Most recently, by coupling the $C=O$ stretching band of the reactant
and the solvent molecules to a Fabry-Perot cavity mode, the chemical reaction rate
of para-nitrophenylacetate and 3-methyl-para-nitrophenylbenzoate has been enhanced\cite{Enhance-DARK-CAVITY-2023}.

The ability to control, i.e., to slow down or enhance chemical reactions, by a seemingly simple setup of reactions inside a cavity made of two parallel mirrors is fascinating.
Unfortunately, currently, theory and experiment have not yet fully converged.
See for example Ref.\citenum{2021-fail-to-reproducibility} where they failed  to reproduce the
enhanced rate of cyanate ion hydrolysis reported by Hiura et al\cite{hiura2021}.
This might be due to the recent findings indicate that
the cavity can introduce friction between the molecules and the solvent. This friction can
enhance the reaction rate when it is low and inhibit it when it becomes significant. See for example study of  the cis−trans isomerization of HONO by Sun and Vendrell\cite{Oriol2022} and references therein. {\color{black} In Ref. \citenum{JOEL-Yuen-Zhou2019resonant}, Yuen-Zhou and his co-workers have shown that the ratio between rate coefficients inside and outside the cavity decreases as the energy of the lower polariton mode decreases (with increasing Rabi splitting), as expected from Marcus's theory (see Fig. 4 in Ref. \citenum{JOEL-Yuen-Zhou2019resonant}).}

The theory presented below is based on the non-Hermitian formalism of quantum mechanics (NHQM), which has previously been utilized in studying the cavity effects in chemistry.
It led to the discovery of novel phenomena, such as the novel mechanism for generating high-purity single-photon emission at high repetition rates from a coherent light source\cite{ANAEL-PRL-2023}, and the conditions enabling the upper bound polariton to penetrate into the continuum and become finite lifetime metastable state\cite{moiseyev2022polariton}.
 NHQM  enables the calculations of the complex discrete poles of the scattering matrix{, i.e., a single NHQM eigenstate instead of a Hermitian wavepacket}.
We will use these poles as a basis set in  calculating cross sections and reaction rates.
The advantage of this non-standard approach is in the ability to isolate the structures in the cross sections from the continuum background.
A single time independent resonance  complex pole may include all measurable information of a specific quantity in spite of the fact that in Hermitian quantum mechanics one needs to carry out wave packet dynamical calculations to describe this phenomena.
 This theoretical approach directs us toward identifying which kinds of collision reactions in gas phase can be \textit{enhanced by conducting them between two mirrors} where the distance between them is carefully chosen.
 Recently we have shown that by using the complex poles as a basis set  a remarkable agreement between theory and experiment was obtained.
 This was shown without using any fitting parameters, where the \textit{only} input parameters for  the cold molecular collision between triplet excited  Helium  and hydrogen molecule where the charges and masses of electrons and nuclei and the Planck constant\cite{deba_3S,deba_3P}. We expect to get similar high agreement between theory and experiment in our study of
collision reactions such as $A+BC \to AB+C$ where a finite lifetime activated complex $[A...B...C]^\#$ is formed and its life time is associated with the complex resonance poles of the scattering matrix.
Notice that these complex poles describe predissociation of metastable  states of the activated complex.
    This is to differ from the isomerization reactions mentioned above\cite{Oriol2022} where the systems is a bound system and the poles do not appear in the complex energy plane but are on the real energy axis. This is an essential point for the theory presented below that provides the conditions for which collision reactions in gas phase (no frictions) can be enhanced in dark cavity due to the quantized field mode states in the vacuum.  \\

Our strategy is outlined as follows:

{First, the principles of the mechanism that enable the enhancement of reactions in the gas phase by {\color{black} quantized radiation field modes} in a cavity are described intuitively without equations using the standard (Hermitian) quantum mechanics formulation. Second,  a brief overview of the fundamental concepts pertinent to this study will be provided.} This will encompass the definition of discrete complex poles within the scattering matrix and its role as a basis set.
In addition, it describes the {predissociation-}resonance metastable  states of the activated complex $[ABC]^\#$, and the nuclear adiabatic theorem that simplified the calculations of  the finite-lifetime resonances of $[ABC]^\#$.
Second, the conditions that should be satisfied in order to enhance the rate of a reaction in a cavity will be derived.
We demonstrate that the reaction rate for $O+D_2\to [ODD]^\# \to OD+D$ can be increased by placing it inside a dark cavity.
The conditions for the enhancement are obtained through tuning the distances between the two cavity-mirrors. {\color{black} Note that cavities with different distances between the two mirrors can have multiple modes resonant with a given molecular transition (see Fig.2d in Ref.\citenum{thomas2024non}). }
{This tuning supports a quantized field mode that couples two resonance states, the occupied state, referred to a transition state  (TS) and an  unpopulated one referred to a dynamical barrier (DB) resonance state.
The DB state remains unpopulated when the distance between the two mirrors is detuned.}
Third, we will  discuss a possible feasible experiments based on the conditions derived herein.
Last we conclude.
{\color{black} In order to convince readers that it is worth the effort to explore a new theoretical concept rather than using standard approaches in the calculation of reaction rates, the results of our nonhermitian theory were introduced before delving into the formalism. This was done to attract the attention of readers to the above findings, which might help experimentalists select reactions whose rates can be enhanced inside the cavity.{\color{black} Notice that the applications of the theory presented here are applied to chemical reactions inside a dark microwave cavity. However, the theory is applicable to any other type of cavity and is not limited to microwave cavities only.}}
\\

{\section{An intuitive (Hermitian) explanation of the cavity-induced reaction rate enhancement}

{Let us assume we carry out standard wavepacket calculations to describe the transition from reactants to products on a single electronic multi-dimensional potential energy surface (PES).
Although in this paper we employ the non-Hermitian formalism, in which this dynamics is described differently using the non-Hermitian time-independent Hamiltonian.
In this Section we do not provide a non-Hermitian  explanation  since the conclusions are the same.
In order to  cross the potential barrier that separates the reactants from the products, the collision energy must be at least equal to the energy of the barrier.
This is because the transition effect due to tunneling is negligible.
However, when the vibrational frequency of the colliding molecules changes substantially during the collision then the transition from reactants to products can be made at a lower collision energy.
{\color{black} This is enabled due to transfer of energy from the translational motion along the reaction coordinate into the lowest bound state of the vibrational mode which is perpendicular to the reaction coordinate}. {\color{black} Notice that the coupling of the molecular modes with the quantized field modes does not always enhance the coupling between the translational coordinate and the orthogonal vibrational coordinates.
The just mentioned translational-vibrational coupling is enhanced only when the frequency of the bound vibrational modes varies non-monotonically at the transition state, which is located near the saddle point where a passage from reactants to products occurs. }
This description is equivalent to  bypassing the static potential barrier along the 1D reaction coordinate but in the  multidimensional potential energy surface (PES).
In the multidimensional PES  dynamics, the transition from reactants to products is through a saddle point, which is the transition state.
When the reactants are far apart the system is described as a delocalized wavepacket, it is a continuum function along the reaction coordinate.
However, around the PES saddle-point the wavepacket  becomes a Gaussian like function (without a node along the coordinates, this important for identifying the transition state).

{\color{black} Thus, the reaction rate depends on the strength of the coupling between the translational motion along the reaction coordinate to the bound vibrational mode perpendicular to the reaction coordinate.}
Above we described how the strength of this coupling is due to the changes in the frequency of the vibration along the reaction coordinate. Inside the cavity the coupling of the molecular modes with the quantized field modes enhance the strength of this coupling and thereby enhance the rate of transition from reactants to products.
Within NHQM it is possible to determine a suitable reaction for accelerating the reaction rate in a dark cavity  without performing simulations of the experiment.
In addition, it allows determination of the optimal distance between the mirrors in the cavity that will enable the optimal coupling described above.}

\section{Brief  overview of the fundamental concepts pertinent to this study}

{ Let us focus here on a collision of atom $A$ with diatom $BC$ (reactants) to form a finite lifetime activated complex $[ABC]^\#$, and then proceeds to yield the products $AB+C$.
Reaction rates\cite{miller1983rate} and its cumulative reaction probabilities\cite{moiseyev1995cumulative} can  be calculated by the standard (Hermitian) quantum scattering theory.
In particular, Ref.\citenum{moiseyev1995cumulative} presents the transition from Lippmann-Schwinger formula for transition probability  to Flux operator, calculated at any given divided surface.
In standard quantum scattering theory the Green operator is defined as $\hat G^+(E)=\lim_{\epsilon\to 0} (E+i\epsilon- \hat H_{ABC})^{-1}$ where $\hat H_{ABC}$  is the  real physical Hamiltonian for the $A+BC\to AB+C$ reaction.
Notice that the term $i\epsilon$ in the Green operator introduces outgoing wave boundary condition into $|\Phi\rangle =\hat G^+(E) |\phi\rangle$ (see solution to exercise 10.1 in Ref.\citenum{NHQM-BOOK}). Thus, the Green operator is a non-Hermitian operator. Let us quote from Ref.\citenum{moiseyev2009feshbach}:
``The branching of quantum mechanics to standard (Hermitian) formalism and non-Hermitian (NH) formalism is associated with the decision to express the exact energy spectrum with one of the two possible self-consistent like problems where the use of the Green operator imposes an outgoing boundary condition on the solutions of the time-independent Schr\"odinger equation.''
Within the framework of the NHQM formalism the Green operator is described using all the discrete complex poles of the scattering matrix.
That is, the NHQM Green operator, in its spectral representation, is defined as $\hat G(E)= \sum_{pol}\frac{|E_{pol})(E_{pol}|}{E-E_{pol}}$, were $E_{pol}$  is a complex number. {\color{black}
Note that here we expand the non-Hermitian Green operator in a basis set of the complex poles of the Hamiltonian, and we should thus use the complex inner product (so-called c-product\cite{NHQM-BOOK}) rather than the scalar product that is commonly used within the hermitian formalism. Namely, $|E_{pol})=|E_{pol}^*\rangle$. }

Not all complex poles in NHQM are linked to resonance states; nevertheless, all resonances are indeed complex poles of the scattering matrix, which we designate as physical complex poles.
These physical and non-physical discrete complex poles can serve as a basis set for calculating reaction rates for structure-less transition probabilities across a potential barrier\cite{Ryaboy-NM-1993-Siegert-Eckart}.
However, in this paper, we only use the physical ones, which are divided into two categories.
The occupied resonance, a transition state (TS), which corresponds to a predissociation of the activated complex, as described above.
{\color{black} The TS resonance is associated with the reaction that takes place outside of the cavity (when the two mirrors are sufficiently far apart) and can be described by a classical mechanism where the reactants have enough energy to overcome the potential barrier of the reaction.  The other physical resonances decay faster than the TS resonance due to  the trapping of reactants for a relatively extended time period due to bound vibrations perpendicular to the reaction coordinate near the saddle point in the potential energy surface (see for example Fig.\ref{CONTOURmap_OD2}). Thus, even in the context of gas phase reactions, the relatively long lifetime of the activated complex arises due to crucial role played by Feshbach resonances, which result from the dynamics in the two dimensional space. To the just discussed effective potential energy barrier that results from the dynamics in 2D space we will refer later as a dynamical potential barrier (DB).  }
As we will explain below, the calculations  of these physical resonances can be simplified by using the nuclear adiabatic approach.
This is done in order to avoid the calculations of the resonance solutions of a multi-dimensional nuclear time-independent Schr\"odinger equation by calculating the resonances of a modified adiabatic one-dimensional (1D) problem.
Since these physical resonances are often embedded inside the 1D potential barrier (of the effective 1D adiabatic potential) we refer to them as dynamical barrier (DB) resonances. For further explanation see Ref.\citenum{Eddy-NM-ArHCL-Mol-Phys-1998}.

The TS predissociation resonance of the activated complex $[ABC]^{\#}$ can be coupled to one of the DB resonances 
with a  quantized field mode.
The TS resonance is the only one associated with the vacuum of the cavity and therefore is the only physical resonance that is occupied before tuning the distance between the two mirrors of the cavity to couple the occupied TS state with one of the selected DB resonances.
{Before delving into the cavity effect on chemical reactions, 
we must first describe the nuclear adiabatic theorem  obtained from the solution of the time-independent Schr\"odinger equation using the  reaction path Hamiltonian.}


In the nuclear adiabatic theorem it is assumed that the motion the products (as they collide to create the activated complex and yield the chemical reaction products) is much slower along the reaction coordinate compared to the motion in the perpendicular directions to the reaction coordinate\cite{pawlak2015adiabatic,pawlak2017adiabatic}.
As we will show below the nuclear adiabatic theory  enables us to simplify the calculations of the predissociation resonance and dynamical barrier states.

\subsection{{The Reaction path Hamiltonian from a potential energy surface}}

 The potential in multi-dimensional space in mass weighted coordinates is given by
    $V(X,Y_1,Y_2,...,Y_{3N-6})$,
where $X$ is the reaction coordinate and $\{Y_j\}_{j=1,2...,3N-6}$ are  perpendicular coordinates to $X$.
 {\color{black} Therefore,
$(X=0,Y=0)$ is a saddle point, which is associated with a transition state as determined by electronic structure calculations.}
 Here,  $N$ is the number of atoms in the $A+B$ (incoming channel), which equals to the number of atoms in $C+D$ (outgoing channel), and $3N-6$ is the number of internal degrees of freedom.
 {However, only $3N-7$ degrees are relevant since one degree corresponds to the reaction direction.
 The frequency of a transition state along the reaction has an imaginary value and for the reactant/products it is zero since it corresponds to the vibration between (minimum) two species that are far apart.
 Therefore, approximately,}
 we can describe this reaction path Hamiltonian (RPH) in the vicinity of the reaction coordinate
where,
\begin{equation}
V(X,Y_1,Y_2,...,Y_{3N-6})\approx V_{r.c}(X)+\sum_{j=1}^{3N-7} \frac{\mu\omega_j^2(X)}{2}Y_j^2 ,
    \label{PES-multidimension}
\end{equation}
where $\mu=m_Am_{BC}/(m_A+m_{BC})$ is the reduces mass and {$V_{r.c.}(X)$, the 1D potential (corresponding to the electronic many-body solutions at different X) calculated along the reaction coordinate supports a potential barrier at $X\equiv 0$.}
{\textit{The predissociation resonance solutions are the eigenfunctions of $\hat H_{RPH}$ when  outgoing wave boundary conditions are imposed. } The narrowest predissociation resonance with energy which is approximately as the energy in classical trajectory calculations  that describe the reaction from reactants to product is associated with the transition state. Here we get into an important issue in our study. Since the TS is associated with the narrowest predissociation resonance all other resonances decay faster. The ability to couple the TS with one of the broader resonance (denoted here as DB resonance) in a dark cavity depends on the ability to have a quantized field mode $\hbar\omega_{cav}=Re[E_{TS}]-Re[E_{DB}]>0$. \textit{This condition  is crucial for enhancement of rates of reaction in a dark cavity. }}

Using the adiabatic approximation the spectrum (bound and continuum states of complex poles when the outgoing boundary conditions are applied) this RPH  is reduced into 1D effective Hamiltonian
$$\Hat H_{ad}=\frac{\hat p_x^2}{2}+V_{ad}(X)$$
where
 \begin{equation}
   V_{ad}(X)=  V_{r.c}(X)+\hbar\sum_{j=1}^{3N-7}  \omega_j(X)(n_j+\frac{1}{2}) \equiv V_{SB}(X)+V_{DB}(X)
     \label{ADIBATIC-PES}
 \end{equation}
 For $n_j=0$ we get the adiabatic Hamiltonian for the ground vibrational states of the normal modes perpendicular to the reaction coordinate.
 In the adiabatic potential given in Eq.\ref{ADIBATIC-PES} there are two type potential barriers. {\color{black}  A static potential barrier (SB), $V_{SB}(X)$, is associated with $V_{r.c.}(X)$.
 The dynamical potential barrier (DB), $V_{DB}(X)$, results from the bound vibrational modes that are perpendicular to the reaction coordinate $X$. The sum of these provides the effective 1D adiabatic potential. }
 The complex poles that will be obtained in the next section are the Feshbach resonances that would be obtained from the study of the dynamics of the multi-dimensional PES as given in Eq.\ref{ADIBATIC-PES}. This approach has been previously used  to study the collisions of hydrogen atom with \color{black}$ArCl$} Van der Waals molecule\cite{Eddy-NM-ArHCL-Mol-Phys-1998} and for several collision  reactions, such as   $H+H_2$\cite{Friedman-Thrular1991CPL,Rice1994resonance}. {\color{black} Notice that the electronic potential surface  does not depend on the mass of the nuclei, and therefore it is the same for any isotope that is used in the experiments. The use of deuterium rather of hydrogen is desirable in order to increase the number of Feshbach type predissociation resonances.   }   It's  important to note that  the complex  physical poles are embedded inside the dynamical potential barrier that is
obtained due the use of the adiabatic approximation described above, and they are \textit{not} embedded in the so-called static potential barrier of $V_{r.c.}(X)$. This is a crucial
point in our derivation of
the conditions where the quantized field modes in a dark cavity are not excited (i.e., in its vacuum field
 state).

 Following our theoretical approach to enhance the reaction rate within the cavity, we must ensure that the dynamical potential barrier supports complex poles which are not obtained (even not approximately) from the static potential barrier. Let us discuss the following reactions
 $H+ArCl \to [ArHCl]^\#\to H+Ar+Cl$ and $O+D_2\to [ODD]^\# \to OD+D$ as  examples for  reactions that can be enhanced by a dark cavity.\\

\subsubsection{The enhancement of the rate of the reaction $H+ArCl \to [ArHCl]^\#\to H+Ar+Cl$}
This reaction has been studied in details in Ref.\citenum{Eddy-NM-ArHCL-Mol-Phys-1998} in free space (when the distance between the two mirrors is taken to be infinitely large). In this work it has been shown that the predissociation resonances of $ [ArHCl]^\#$ which results from the vibrations perpendicular to the reaction coordinate (when hydrogen atom oscillates between Ar and Cl) are accurately obtained from the 1D effective potential barrier that is obtained within the framework of the nuclear adiabatic approximation. Here we wish to show that in this case where all complex poles are associated with the dynamical potential barrier and are physical poles the rate of the reaction is enhanced by the dark cavity that couples the TS resonance with one of the DB resonance of $ArHCl$. {\color{black}  The physical poles are associated with the Feshbach resonances of the 2D potential energy surface, and their associated energies (the real part of the complex energy of the pole of the scattering matrix) are larger than the threshold energy of the 1D effective potential. All other poles are non-physical and can only be used as part of the discrete basis set constructed from all complex poles of the scattering matrix. A simple distinction between the physical and non-physical complex poles can be made by comparing the complex poles associated with shape-type resonances obtained from the 1D potential as a function of the reaction coordinate against the complex poles of the adiabatic effective 1D potential energy curve. The complex poles of the adiabatic 1D potential energy curve approximate the 2D Feshbach resonances mentioned above. The poles associated with the 1D shape-type resonances are false resonances and they are very different from the Feshbach resonances obtained from the 2D reaction path Hamiltonian, which are considered the physical complex poles.}

In Fig.\ref{POT-ArHCl} we show the dynamical potential barrier for the reaction
$$H+ArCl \to [ArHCl]^\#\to H+Ar+Cl$$
and the positions of the TS and DB resonances.
{The potential barrier is symmetric although the reaction is asymmetric (products not equal to reactants) due to the weak van der Waals bonds of ArCl whereas the bonds in the activated complex $[ArHCl]^\#$ are stronger, which results in negligible static potential barrier along the reaction coordinate in comparison to the dynamical potential barriers.}
\begin{figure}[h!]
     \centering
     \hspace*{+0.9cm} 
     \includegraphics[angle=000,scale=0.5]{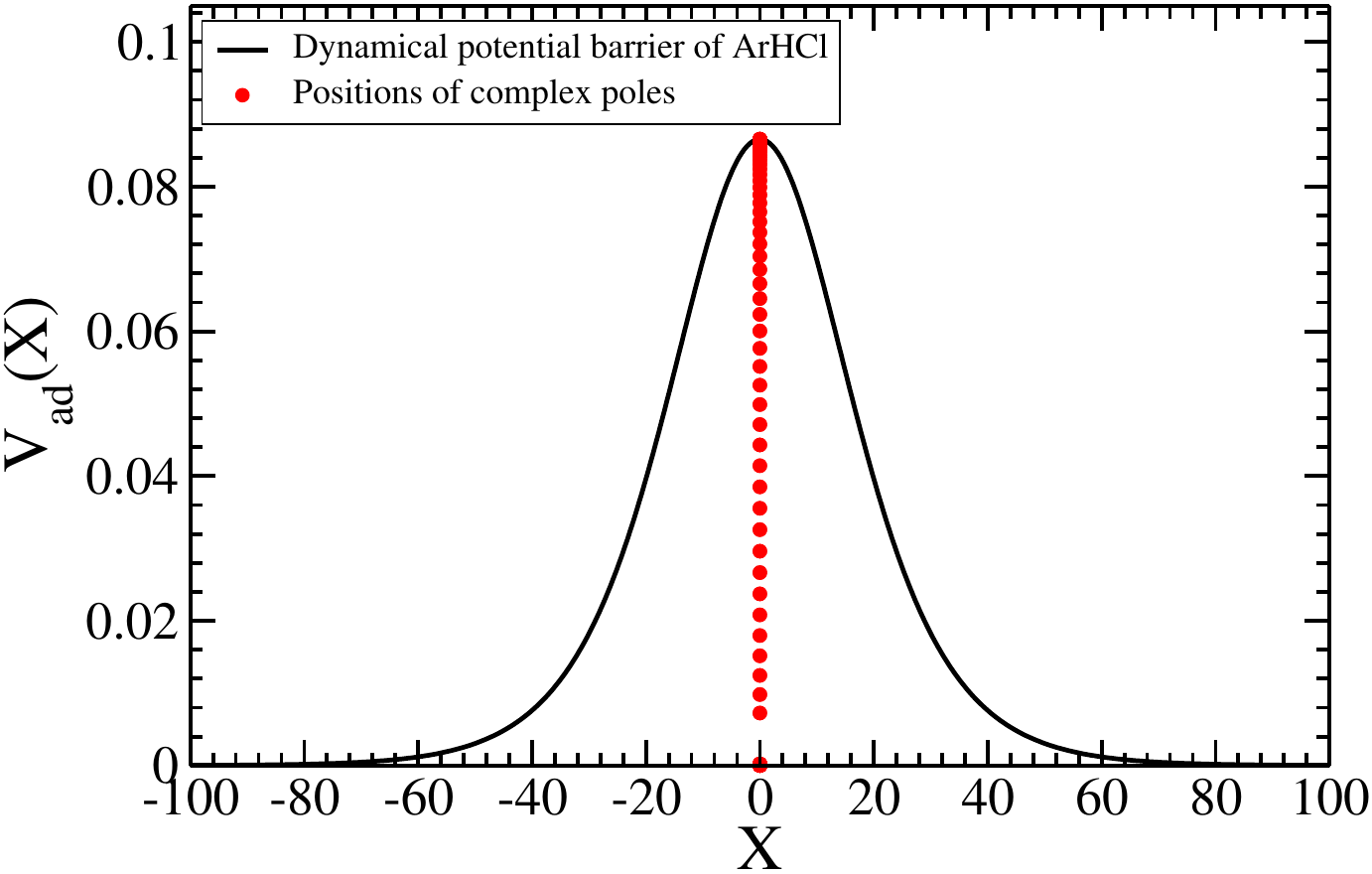}
      \caption{The dynamical potential barrier  (black line) and the positions (red dots) of the complex poles for ArHCl.       The potential is adopted from Ref.\citenum{Eddy-NM-ArHCL-Mol-Phys-1998} and the resonance positions are calculated herein using complex scaling.
      Potential in Hartree and coordinate { also in atomic units}. These resonances are actually the dynamical resonances as obtained from the full multi-dimensional calculations, notice that the static potential barrier along the reaction coordinate is negligible (see Ref.\citenum{Eddy-NM-ArHCL-Mol-Phys-1998}).}
    \label{POT-ArHCl}
    \end{figure}
In Fig.\ref{CPOLES-ArHCl} the predissociation resonances of $ArHCl$ are presented counted by the number of their nodes. The TS and the DB resonance that would be coupled by the dark cavity are marked by a blue color.
{The TS resonance state is localized at the top of the potential barrier, it has the longest lifetime and it looks similar to the Gaussian wavefunction as obtained for a parabolic potential barrier.~\cite{miller1976}}
  \begin{figure}[h!]
     \centering
     \hspace*{+0.9cm} 
     \includegraphics[angle=000,scale=0.5]{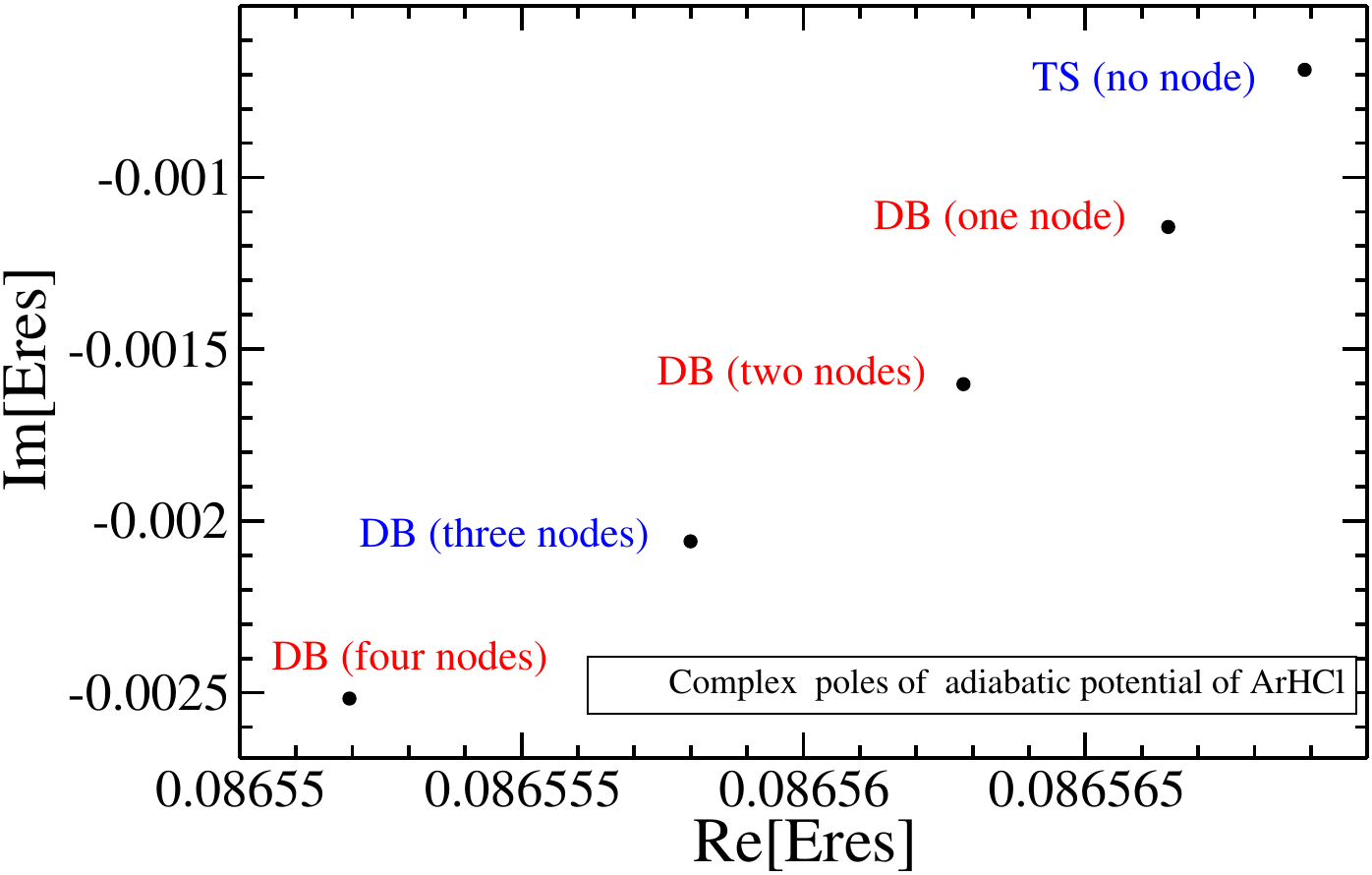}
      \caption{{The highest five resonance complex poles of $ArHCl$, calculated by the uniform complex scaling method.  All of them results from the vibrational states of $ArHCl$  when along the reaction coordinate $H$  has almost a free translational motion with respect to the $ArCl$ center of mass. The number of nodes were counted from the plots of the square integrable complex scaled $|\Psi_{res}(X)|^2$. Notice that the transition state is associated the longest lifetime resonance that is localized at the top of the potential barrier. The cavity is tuned to couple the two states marked in blue. Energies in Hartree.}} 
    \label{CPOLES-ArHCl}
    \end{figure}

Below in Fig.\ref{GammaPOLARITON-ArHCl} we show the enhancement of the rate of this reaction 
when the distance between the two mirrors in the Fabry Perot dark cavity are tuned to have a quantized field mode with the frequency $\omega_{cav}=0.195 meV$.
{This quantized field mode corresponds to the resonance condition that couples the TS resonance and the DB resonance, which are  marked in blue in Fig.\ref{CPOLES-ArHCl}.}

{\color{black} This result encourages experimentalists to study such phenomena in the context of either cold or room temperature molecular collisions in gas phase, as for example in Narevicius Lab\cite{Eddy-Lab}. Generally speaking, the relevant temperatures should be high enough as to allow overcoming classically the static potential barrier. In the particular case discussed here (the collision of deuterium diatomic molecules with oxygen)
 the experiment has to be done at room temperature. On the other hand, an experiment involving $ArHCl$ should be done preferably at cold temperatures, because the static potential barrier along the reaction coordinate is negligible.
 In the experiments of Ref.\citenum{Eddy-Lab}, controlling the temperature implies to increase the angle between the angle of the two colliding atomic/molecular beams (this is equivalent to increase the temperature of the collision). }

 \begin{figure}[h!]
     \centering
     \hspace*{+0.9cm} 
     \includegraphics[angle=000,scale=0.5]{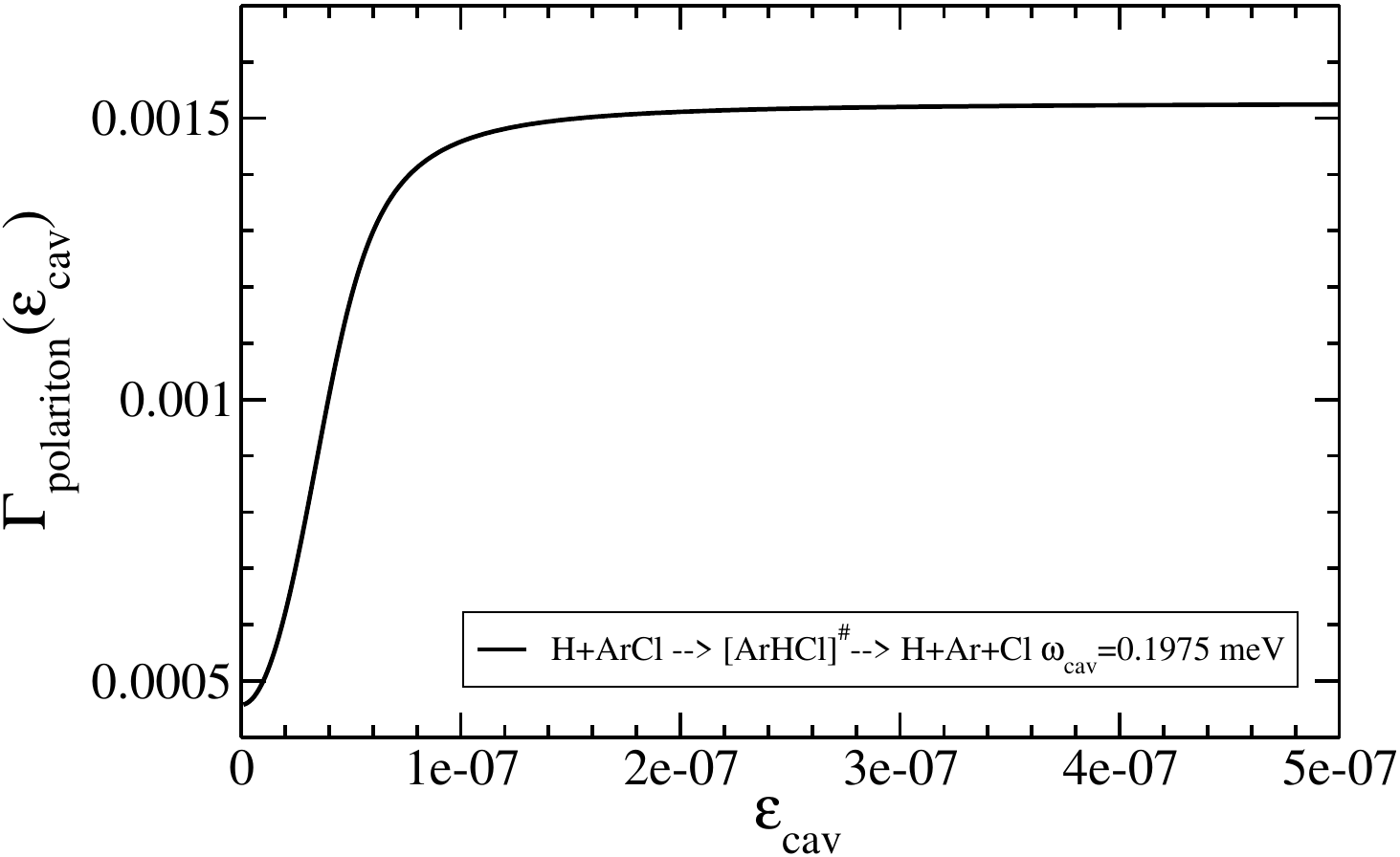}
      \caption{The rate of the reaction $H+ArCl\to H+Ar+Cl$ as a function of the coupling strength between the dark cavity and the $ArHCl$ molecule at resonance condition between the states marked in blue in Fig.\ref{CPOLES-ArHCl}. The cavity can be tuned into this resonance condition by adjusting the distance between the two Fabry-Perot mirrors (see Eq.~\ref{epsilon_VOL_cav}). {Detailed of the calculation are given in Section~\ref{eqss},
      in particular, Eq.~\ref{minGAMMApol}. Rate in Hartree and coupling in atomic units.} {\color{black} Notice that $E_{TS}-E_{DB}(3nodes)\approx \hbar \omega_{cav}(1_{photon}+\frac{1}{2})$}. {\color{black}  Here we used such a distance between the two mirrors that generates one mode resonant with the molecular transition (i.e., $n_{photon}=1$) as mentioned in Ref.\citenum{thomas2024non}.
      The enhancement is saturated quickly, since in this case the probabilities of the upper and  lower polaritons to populate the transition state become about equal as $\epsilon_{cav}$ is increased.
 Roughly speaking, the rate of reaction is about equal to the averaged value of the TS decay rate outside of the cavity and the decay rate of the broad three mode resonance (shown in Fig.\ref{CPOLES-ArHCl}). This point will be explained later by Eq.\ref{GAMMApol}.}}
    \label{GammaPOLARITON-ArHCl}
    \end{figure}

\subsubsection{The enhancement of the rate of the reaction $O+D_2 \to [ODD]^\#\to OD+D$}
\label{xxx}

This reaction represents the most common situation where the chemical bonds in the reactants and the products are stronger than the chemical bonds of the activated complex. In the type of reactions the static potential barrier is much more pronounced than the dynamical potential barriers.
The physical predissociation resonances of the activated complex are associated withe complex poles that are obtained from the 1D effective potential obtained within the framework of the adiabatic approximation where the vibrations perpendicular to the reaction coordinate are taken into consideration.

  In Ref.\citenum{johnson1977classical} Johnson and Winter studied the the effect of vibrational
energy on the reaction of molecular hydrogen with atomic
oxygen. They calculated the  the contour map for the $O+H_2 \xrightarrow[]{\Gamma(\epsilon_{cav}=0)} OH+H $, in which $O$ collides with $H_2$. See below their contour map plot as presented in Fig.1 in their paper.  Notice that in the electronic structure calculations the isotopic effects are not taken into consideration and therefore we can replace hydrogen molecule by deuterium diatomic molecule.
 The product is $OD+D$ and the frequency of the vibration which is perpendicular to the reaction coordinate is denoted by $\Omega_{vib}(X)$. The incoming oxygen atoms collide with deuterium diatomic molecules when they are temporarily trapped for a long time around the saddle point due to the oscillations perpendicular to the reaction coordinate. This temporarily trapping results in the formation of a long lived transition state (TS) predissociation resonance. Nevertheless, there are other excited predissociation resonances which are not populated but, as we will show here, will interact with the TS resonance in a dark cavity resulting in enhancement of the reaction. The different of the masses of the isotopes will be taken into consideration in the next step where we will describe potential energy surface for this reaction using the reaction path Hamiltonian representation.
 \begin{figure}[h!]
     \centering
      \hspace*{+0.9cm} 
      \includegraphics[angle=000,scale=0.6]{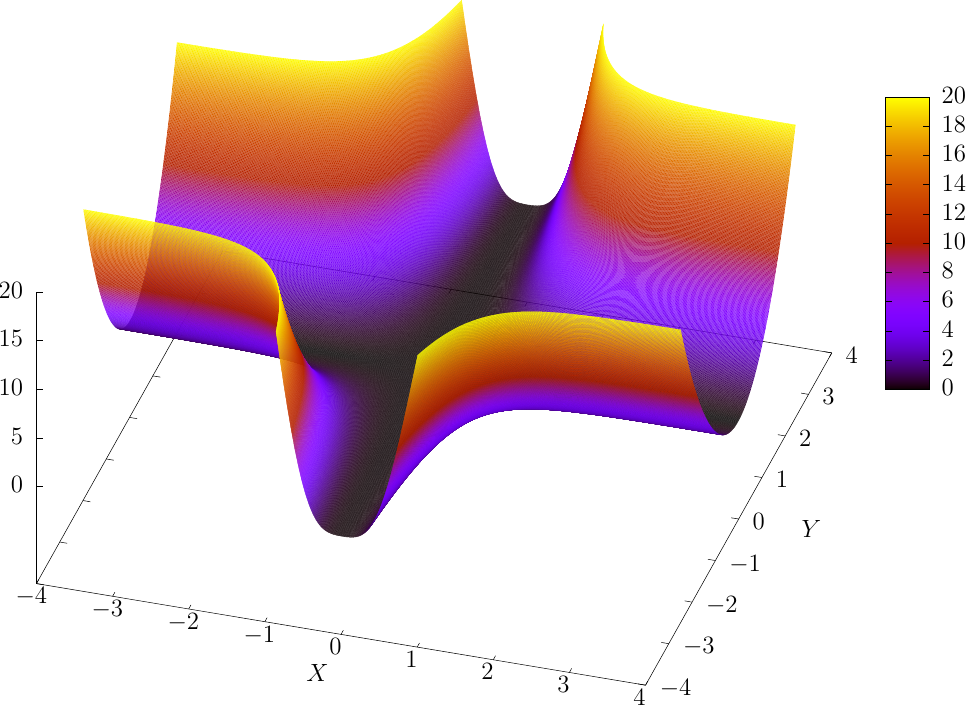}
      \caption{ { A 3D plot showing the potential energy
      for the reaction where oxygen ($O$) collides with a deuterium ($D_2$) molecule to form $OD$ and $D$. The reaction follows a path (so called reaction coordinate)  that goes over a high-energy point called the saddle point, forming a temporary structure of an activated complex labeled $[ODD]^{\#}$. The Y-axis is perpendicular to the direction of the reaction rate coordinate, X. Both axes are measured in Bohr.
      }}
    \label{3D-OD2-PES_OD2}
    \end{figure}

    \begin{figure}[h!]
     \centering
     \hspace*{+0.9cm} 
     \includegraphics[angle=000,scale=0.6]{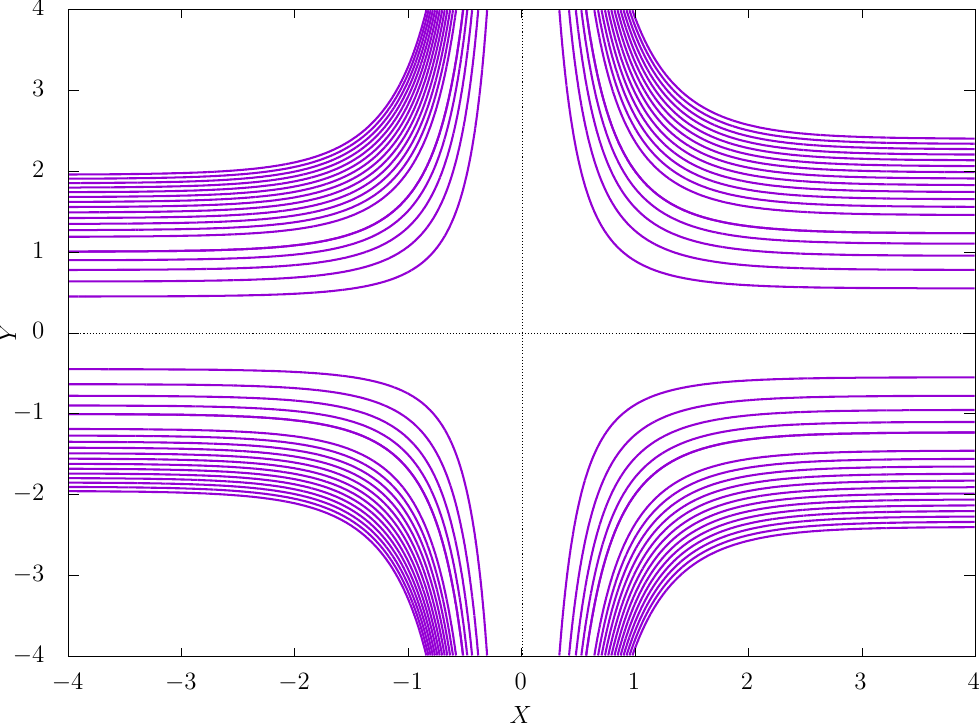}
\caption{  A 2D contour map for the reaction where oxygen ($O$) collides with a deuterium ($D_2$) molecule to form $OD$ and $D$. As in the 3D plot of the potential surface the focus here is on the weaker, "soft" vibrations along the Y-axis, compared to the vibrations in the original reactants ($O + D_2$) and the products (OD + D). }
    \label{CONTOURmap_OD2}
    \end{figure}

   \begin{figure}[h!]
     \centering
     \hspace*{+0.9cm} 
      \includegraphics[angle=000,scale=0.6]{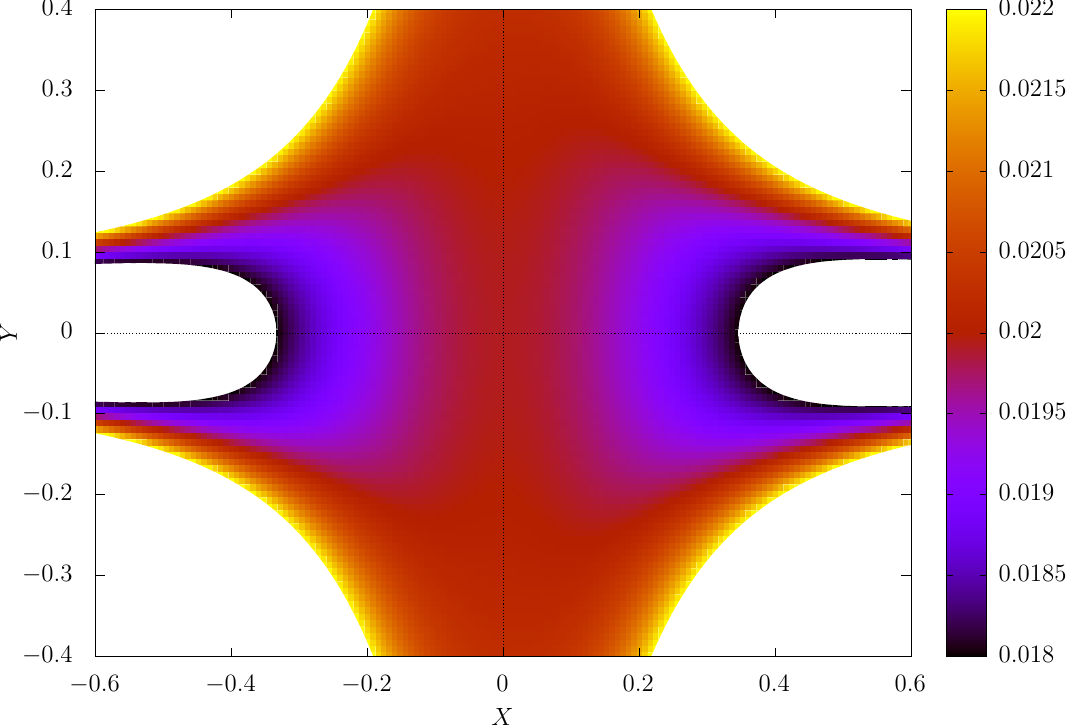}
\caption{  A zoomed-in colored 2D contour map as shown in Fig.\ref{CONTOURmap_OD2}, here one can see the static potential barrier unlike in Fig.\ref{3D-OD2-PES_OD2} and \ref{CONTOURmap_OD2}. }
    \label{CONTOURmap_OD2-SB}
    \end{figure}

 The 3D and 2D potential energy surfaces as described in Fig.\ref{3D-OD2-PES_OD2} and Fig.\ref{CONTOURmap_OD2} were obtained respectively
 by using Eq.\ref{PES-multidimension} and Eq.\ref{ADIBATIC-PES}.
 The information for this plot comes from Johnson and Winter in 1977, see Ref.\citenum{johnson1977classical}.
 On this scale, one can not see the energy barrier in the X-direction.
 Our focus is on the weaker, "soft" vibrations along the Y-axis, compared to the vibrations in the original reactants ($O + D_2$) and the products ($OD + D$).
 In Fig.\ref{CONTOURmap_OD2-SB} one can see the static potential barrier along the reaction coordinate.
 As explained in Section(II)  the reactants $O$ and $D_2$ approaching one another as they move along $X<0$.
 Due to the relative high frequency of  the ground vibrational level of $D_2$ the collision energy is about equal to the height of the static potential barrier.
 However,  as they get to be close enough to form the activated complex $[ODD]^\#$ the energy transfer from the excited vibrational soft mode of $[ODD]^\#$ to the X-component of the translational kinetic energy will enable to bypass the potential barrier. }
 Notice that $V_{SB}(X)$ is approximately described as
an asymmetric Eckart potential barrier given by (Hartree unites) as a
function pf the reaction coordinate given in Bohr unites,
\begin{equation}
    V_{SB}(X)=-0.0180244\cosh^2(0.05)[\tanh(X+0.05)-
    \tanh(0.05)]^2+0.0180244 \times e^{2\cdot 0.05}
    \label{eqVSB}
\end{equation}

\begin{figure}[h!]
     \centering
     \hspace*{+0.9cm} 
     \includegraphics[angle=000,scale=0.5]{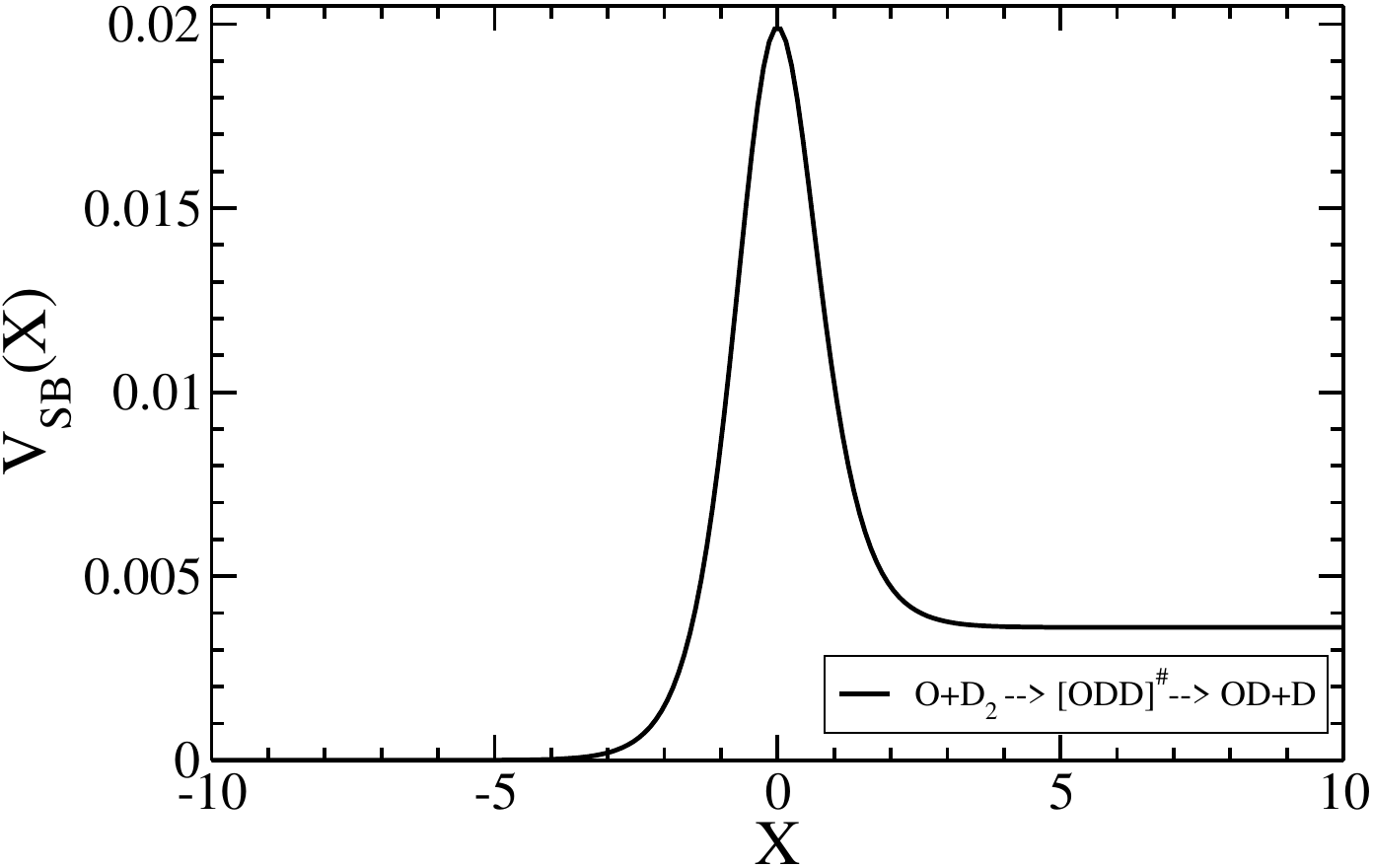}
      \caption{ Static potential barrier as obtained from the 1D electronic structure calculations along the reaction coordinate  {(see Eq.\ref{eqVSB} in the text)}. {\color{black} Potential in Hartree and coordinate in Bohr. Notice that the energy of the reactants, $O+D_2$, is taken as zero.}}
    \label{figVSB}
    \end{figure}

The static potential barrier $V_{SB}(X)$ is shown in Fig.\ref{figVSB}

\begin{figure}[h!]
     \centering
     \hspace*{+0.9cm} 
     \includegraphics[angle=000,scale=0.5]{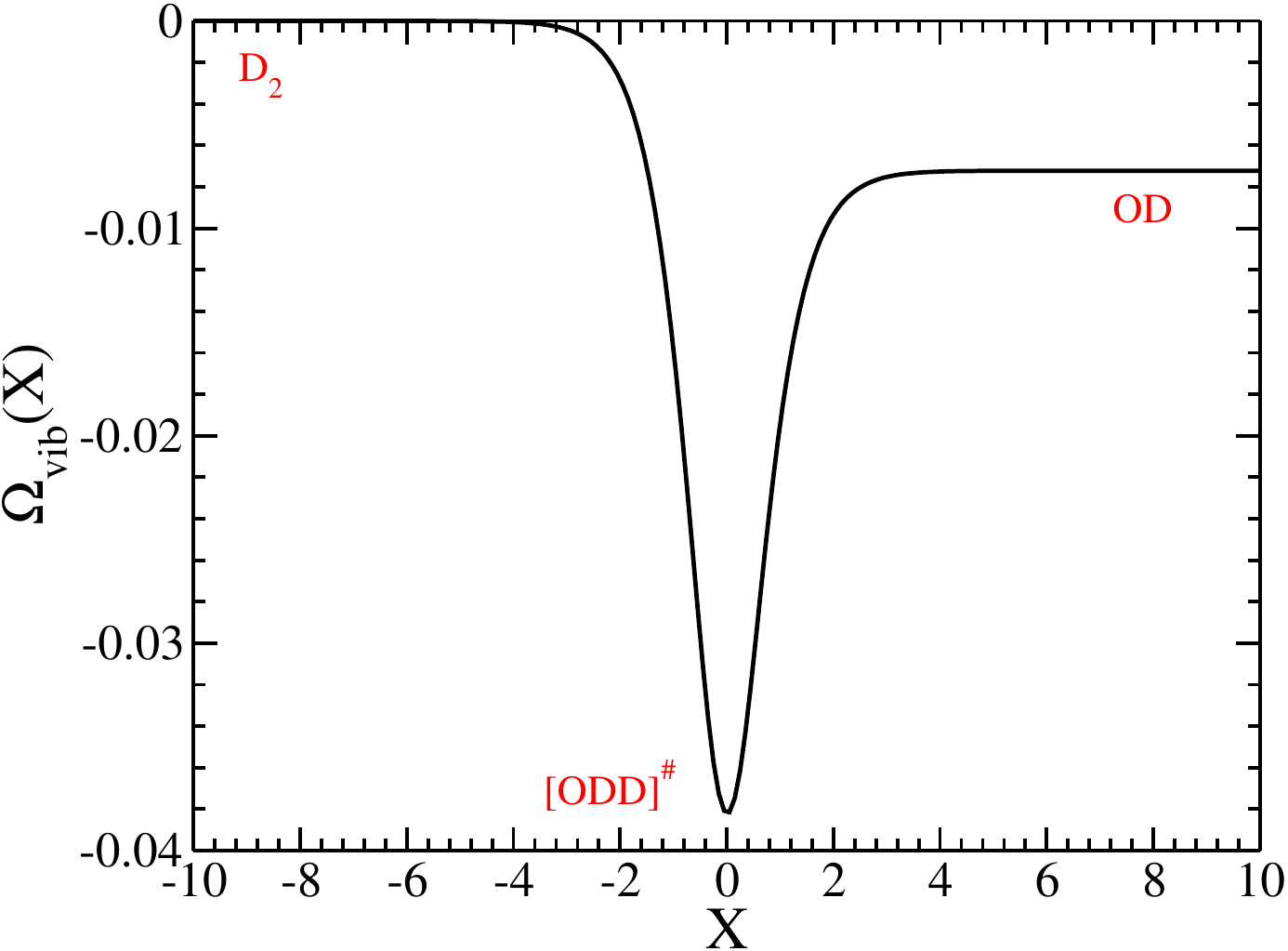}
      \caption{ The $O+D_2 \xrightarrow[]{\Gamma(\epsilon_{cav}=0)} OD+D $ reaction. The frequency of the vibration perpendicular (Y)  to the reaction coordinate  as a function of the reaction coordinate X. Notice that the  vibrational frequency of the reactant $D_2$ is taken as a reference and therefore the frequency of the product  $OD$ gets a negative value. Frequencies in Hartree and coordinate in Bohr.}
    \label{OMEGA_OD2}
    \end{figure}

The $V_{DB}(X)=\Omega_{vib}(X)(n_Y=0+1/2)$  is given by $V_{DB}(X)=\Omega_{vib}(X)(n_{vib}=0+1/2)$ where $\Omega_{vib}(X)$ is described in Fig.\ref{OMEGA_OD2}.

\newpage

\begin{figure}[h!]
     \centering
     \hspace*{+0.9cm} 
     \includegraphics[angle=000,scale=0.5]{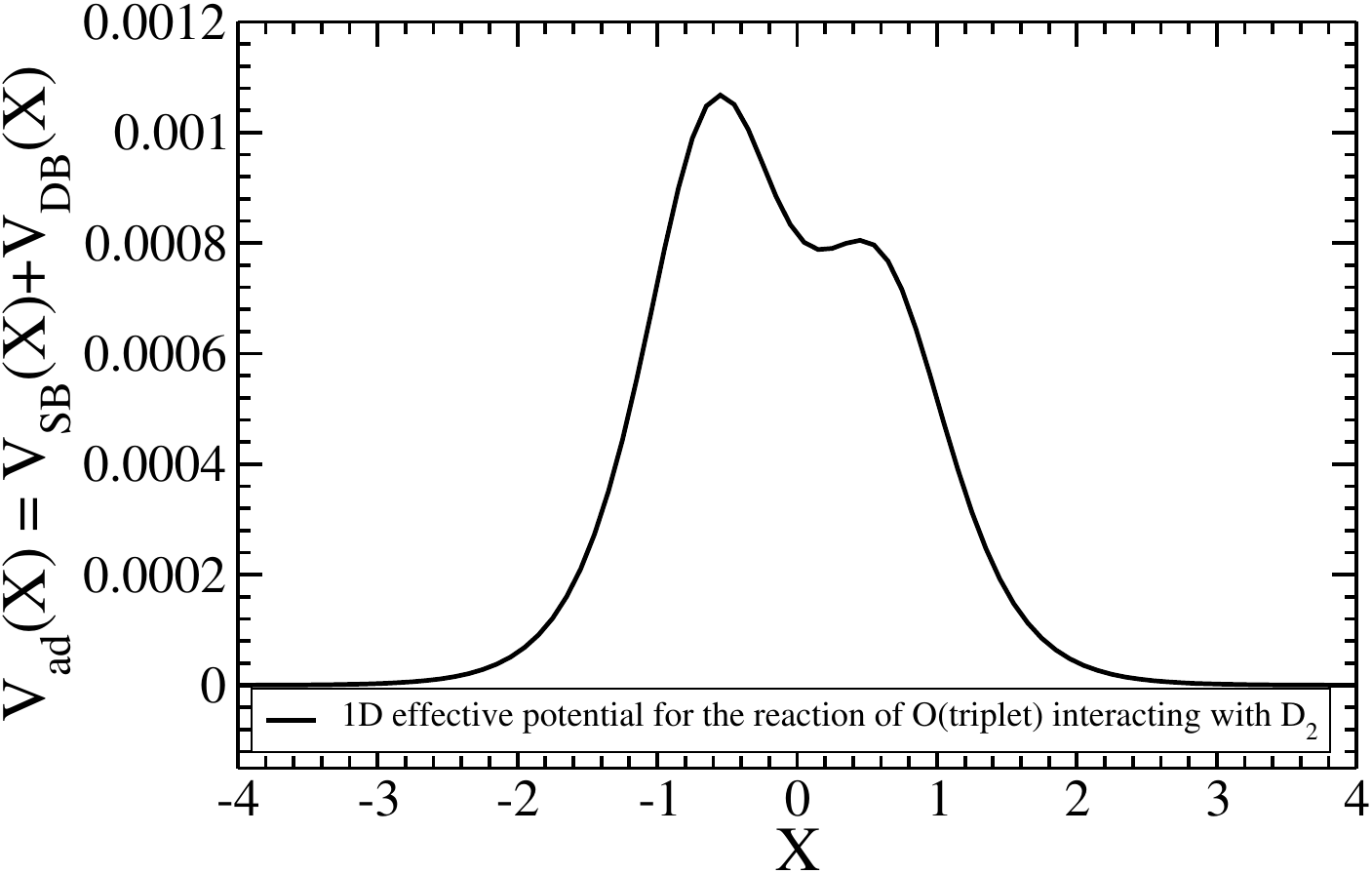}
      \caption{ The adiabatic potential, obtained by adding the static potential and the dynamical barrier potential as described in the text. This adiabatic potential is exactly equal to the potential given in Eq.6.3 and in Fig.1c in Ref.\citenum{friedman1997barrier}   for the $O+D_2 \to  OD+D $ reaction.  Potential in Hartree and coordinate in atomic unites. }
    \label{Vad_OD2}
    \end{figure}

The sum $V_{SB}(X)+V_{DB}(X)$ provides the adiabatic potential that includes the effect of vibrational
energy on the reaction of molecular hydrogen/deuterium  with atomic
oxygen, $V_{ad}^{O+D_2\to OD+D}(X)$ which is given in Fig.\ref{Vad_OD2}.
 Notice that the height of the  adiabatic potential barrier  
is about 0.0015 Hartree, more then order of magnitude smaller in comparison to the height of the static potential barrier of 0.02 Hartree as shown in Fig.\ref{figVSB}.
It shows how the coupling between the vibrations perpendicular to the reaction coordinate to the kinetic energy component along the reaction coordinate enables the reactants to by passing the static potential barrier to  get the reaction products $OD+D$.

\newpage

\begin{figure}[h!]
     \centering
     \hspace*{+0.9cm} 
     \includegraphics[angle=000,scale=0.5]{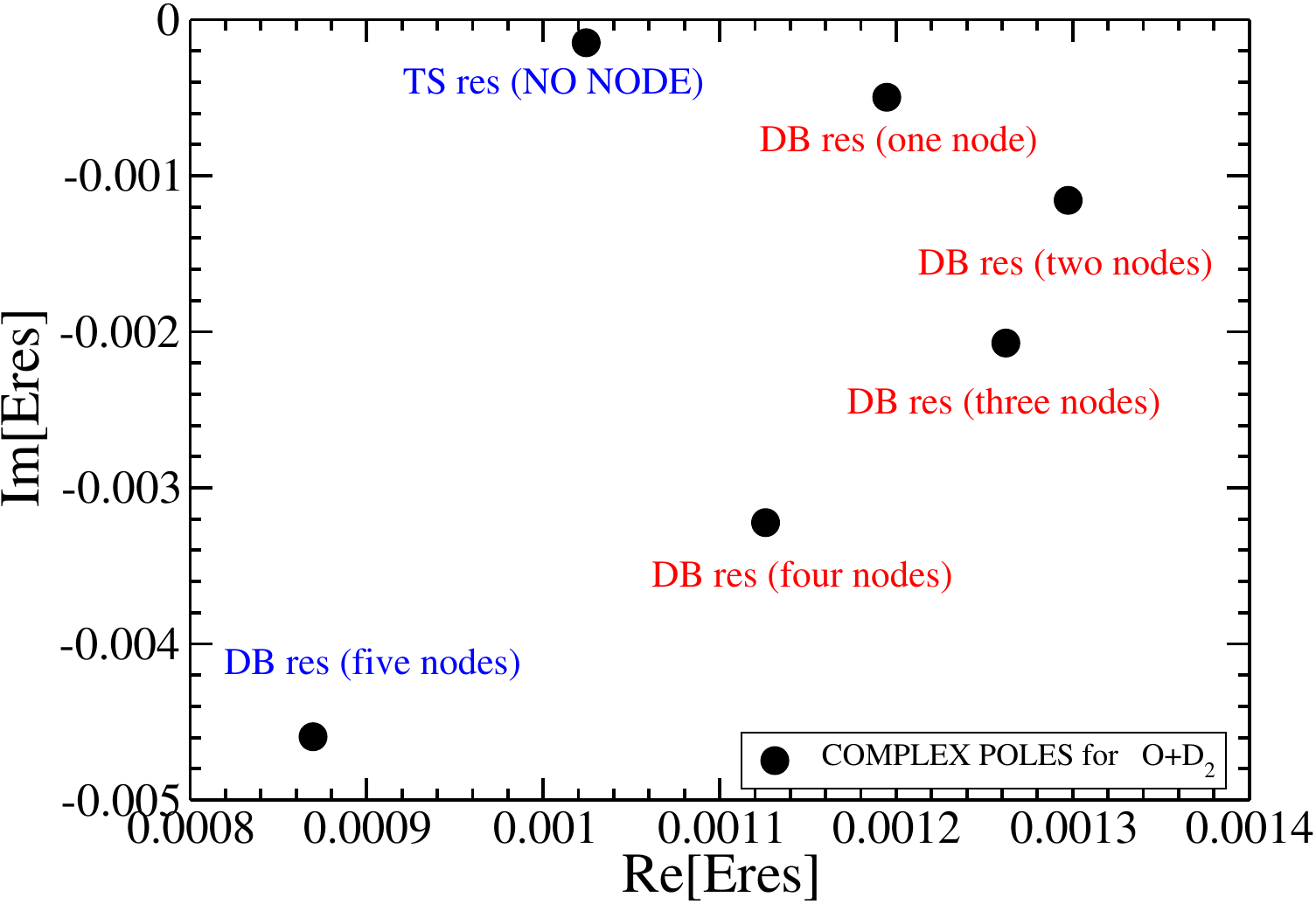}
      \caption{ {The highest six complex poles of $[ODD]^\#$, calculated by the uniform complex scaling method. Energies in Hartree. The two resonances that will be coupled by the cavity are marked in blue. }}
    \label{POLES_OD2}
    \end{figure}

By using uniform complex scaling, where the X coordinate was rotated into the complex plane by an angle of  $\theta=0.75 rad$ \cite{NHQM-BOOK}, the resonances for the adiabatic Hamiltonian for $OD_2$
  were calculated and counted by the number of their nodes. The results for the transition state (TS) and dynamical barrier (DB) resonances are given in Fig. \ref{POLES_OD2}. By comparing the static and adiabatic potentials presented in Fig. \ref{figVSB} and Fig. \ref{Vad_OD2}, it is clear that the complex poles in Fig. \ref{POLES_OD2} are all associated with the dynamical barrier potential and not with the static potential barrier. Notice that the transition state is associated with the longest lifetime resonance, which is localized at the top of the adiabatic potential barrier (approximately 0.0015 hartree). The other resonances (denoted as dynamical potential-barrier resonances) are associated with the excited vibrations that are perpendicular to the reaction coordinate. The cavity is tuned to couple the two states marked in blue in Fig. \ref{POLES_OD2}.

The ability to enhance the rate of the reaction $O+D_2\to OD+D$ is due to the fact that there is a DB resonance (mark by a blue color) which its position is lower than the position of the TS resonance (mark also by a blue color).
Therefore through  emission of one photon  the TS resonance  will have the same energy as of the five nodes excited DB resonance.
As we will show below in the next section the rate of this reaction is enhanced when the distance between the two mirrors is tuned to have a quantized field mode state with the energy $\hbar\omega_{cav}$ which is equal to the energy gap between the TS and DB resonances.
 That is, $\hbar\omega_{cav}= Re[Eres(TS)-Eres(DB)]=0.0001547 Hartree$.
The condition that  $Re[Eres(TS)-Eres(DB)]>0$  implies that due to the emission of one photon  the two resonances  becomes almost degenerate with the energy of the DB resonance (this resembles the situation in standard stimulated emission).
In the next section we will describe how the two molecular states are mixed with the { quantized radiation field modes} to form upper and lower polaritons.
This condition should be satisfied in any other reaction that its rate can be enhanced by a \textit{dark} cavity.


\section{The enhancing of the reaction rate  inside a dark cavity : theory and example} 

\label{eqss}

In scattering calculations, as in the study of chemical reactions, it is preferable to use the acceleration {gauge}  representation since the dipole transition matrix elements vanish at the asymptotes. Within the acceleration gauge representation\cite{moiseyev-sindelka-2022AG} the cavity induces a dipole interaction between the poles given by $0.5\alpha_{cav}\langle TS|\partial V_{ad}/\partial X|DB\rangle$, where $\alpha_{cav}=\epsilon_{cav}/\omega_{cav}^2$ (see Ref.\citenum{moiseyev-sindelka-2022AG}) are  the cavity parameters.
{\color{black}  Therefore, in the acceleration gauge representation
\begin{eqnarray}
\hat H_{polariton} =&&\nonumber \\ &&[\hat T_X+\hat T_Y+V(X,Y)]|0_{photon}\rangle\langle 0_{photon}|+ \nonumber \\ &&[\hat T_X+\hat T_Y)+V(X,Y)+\hbar\omega_{cav}]|1_{photon}\rangle\langle 1_{photon}|+ \nonumber \\ && \frac{\alpha_0}{2}\frac{\partial V(X,Y)}{\partial X}(|0_{photon}\rangle\langle 1_{photon}|+|1_{photon}\rangle\langle 0_{photon}|)
\label{POL-HAM-AG}
\end{eqnarray}
   where {\color{black} $V(X,Y)$ is the potential as obtained from the RPH}, $\alpha_0=\epsilon_{cav}/\omega_{cav}^2$, $X,Y$ are the mass weighted coordinates, and $T_X=-(1/2)\partial_{XX}$, $T_Y=-(1/2)\partial_{YY} = -(1/2) \sum_{j=1}^{3N-7} \partial_{Y_{j}Y_{j}}$.
For the sake of clarity, let us point out that the Hamiltonian of Eq.~\ref{POL-HAM-AG} corresponds to the standard reaction path Hamiltonian as
derived in Ref.\citenum{miller1980reaction}. Here $X$ stands for the reaction coordinate. The used coordinate system $(X,Y)$ is curvilinear, and this implies a nontrivial form of the kinetic energy operator. In Eq.~\ref{POL-HAM-AG} we have however used a simplified approximated version of the kinetic energy operator, which neglects the Coriolis effects. Such an approximation is justified provided that the reaction takes place in close vicinity of the saddle point. }

   Within the adiabatic approximation the polariton Hamiltonian is given by,
\begin{eqnarray}
\hat H^{ad}_{polariton} =&&\nonumber \\ &&[-\frac{1}{2}\partial_{XX}+V_{r.c.}(X)+\hbar\Omega(X)/2]|0_{photon}\rangle\langle 0_{photon}|+ \nonumber \\ &&[\hat T_X+V_{r.c.}(X)+\hbar\Omega(X)/2+\hbar\omega_{cav}]|1_{photon}\rangle\langle 1_{photon}|+ \nonumber \\ && \frac{\alpha_0}{2}\frac{\partial V(X,Y)}{\partial X}(|0_{photon}\rangle\langle 1_{photon}|+|1_{photon}\rangle\langle 0_{photon}|)
\label{POL-HAM-AD-AG}
\end{eqnarray}
{\color{black}An analogous kind of adiabatic approximation has been used before in the context of the reaction path Hamiltonians e.g. in Ref.\citenum{DC9878400427} (see Eq.~10 therein).}
The TS and DB resonances are obtained by solving Eq.\ref{POL-HAM-AD-AG} by using the complex scaling approach\cite{NHQM-BOOK}.

   Here we are coming to the association of $\alpha_0$ with the length of the distance between the two mirrors denoted by $L$.
   \begin{equation}
       \epsilon_{cav}\equiv \sqrt{2\pi\hbar\omega_{cav}/LA}
   \label{epsilon_VOL_cav}
   \end{equation}
   where A is the area of the mirrors and $LA$ is the volume of the cavity.
   It is clear that when $L\to \infty$ the strength of the
   coupling between the molecules and the cavity is reduced to
   zero.  When $L=\lambda/2$ where $\omega_{cav}=4\pi c/L$ we get
   that $L=4\pi c/\omega_{cav}$. Only  when $\hbar\omega_{cav}$ is
   equal to the difference between the TS energy and the  energy
   of the selected DB resonance   the rate of the reaction will
   be enhanced by the cavity as we will show here. {\color{black} Notice that, although we discuss here the interaction of a single molecule with the cavity, our presented  theory holds also when there are
$N$ identical molecules inside the cavity that do not interact with each other (but all of them interact with the same dark cavity mode). In such a case, $\epsilon_{cav}$
for a single molecule is multiplied by $\sqrt{N}$ (see Ref.\citenum{herrera2016cavity}).}

  A $2\times 2$ complex symmetric matrix representation of Eq.\ref{POL-HAM-AG} in the field free resonance basis set is given by,
\begin{eqnarray}
&&{\bf H^{pol}_{Mol/Cav}}(\epsilon_{cav},\omega_{cav}) = \nonumber \\ &&
\label{polaritonHAM}
   \begin{pmatrix}
   E_{TS}-i\Gamma_{TS}/2 & \alpha_{cav} d_{[TS],[DB]} \\
   \alpha_{cav} d_{[DB],[TS]} & E_{DB}+\hbar\omega_{cav} -i\Gamma_{DB}/2
\end{pmatrix},
\end{eqnarray}
Notice that, $E_{DB}+\hbar\omega_{cav}=E_{TS}$.
In dark cavity where the energy of TS resonance is larger than the energy of the DB resonance it implies that the TS emits a photon with the energy $\hbar\omega_{cav}$.
The coupling between them is induced by the cavity quantized field modes to create a superposition of the two polaritons where the decay rate of the TS resonance is increased while the rate of the DB resonance is reduced.
This mechanism is described in more details below.

{\subsection{Theory - the cavity effect on the rate of a reaction}}

The interaction of the activated complex in its transition state configuration with { quantized radiation field modes} of a planar cavity is described by a dressed Hamiltonian. We use as a basis set to construct the dressed Hamiltonian  a mixture of molecular and quantized field mode states (so called polaritons).
The minimal basis set for the AC outside of the cavity is constructed from the Siegert complex solutions described in the previous paragraph, see for example in Ref.\citenum{klaiman2010absolute}.  We use for the study of the  reactions inside the cavity two polariton basis functions.
One polariton basis function which describes the initial state
(TS outside of the cavity) is $|TS \rangle |0_{photon}\rangle$. The second basis function is
$|TS\rangle|1_{photon}\rangle$ which is coupled to the initial state by the cavity. {\color{black} The
$|TS\rangle$ state is also metastable (a Siegert type eigensolution) if the activated complex is associated with the
complex eigenvalue $E_{TS}-i\Gamma_{TS}/2$}.
{Therefore, on resonance condition where $$E_{TS}=E_{DB}+ \hbar\omega_{cav}$$ (i.e., due to the coupling between the molecules modes and quantized field mode a  photon is emitted)} and
$$\Gamma_{TS} \ll \Gamma_{DB}$$ (as shown Fig.~\ref{POLES_OD2}), the
polaritonic Hamiltonian in the length gauge (neglecting the self energy term which is proportional to $\epsilon^2$) is given  by,
\begin{eqnarray}
\label{POL-HAM}
&&\hat H_{polarition}=(E_{TS}-i\Gamma_{TS}/2) |TS \rangle |0_{photon}\rangle \langle TS|\langle 0_{photon}| \nonumber \\ &&+
(E_{DB}-i\Gamma_{DB}/2+\hbar\omega_{cav}) |TS \rangle |1_{photon}\rangle \langle TS|\langle 1_{photon}|\nonumber \\ &&+ \epsilon d_{[TS],[DB]}
(|TS \rangle |0_{photon}\rangle \langle DB|\langle 1_{photon}|+|DB \rangle |1_{photon}\rangle \langle TS|\langle 0_{photon}|)
\end{eqnarray}
Often the Hamiltonian of the polaritons is written within the length { gauge} representation. However, in scattering calculations, as in the study of chemical reactions, it is preferable to use the acceleration { gauge} representation since the dipole transition matrix elements vanish at the asymptotes.See Eq.\ref{POL-HAM-AG} for the Hamiltonian for polaritons  in the acceleration {\color{black} gauge} representation.
  A $2\times 2$ complex symmetric matrix representation of Eq.\ref{POL-HAM-AG} in the field free resonance basis set is given by Eq.~\ref{polaritonHAM}.
Notice that, $E_{DB}+\hbar\omega_{cav}=E_{TS}$, in addition, in scattering calculations it is preferable to use the acceleration gauge representation (see for example Ref.\citenum{ABA-PRL}).


In order to study the effect of  increasing the strength of the coupling between the two metastable (resonance) states that are associated with the TS ans the DB state we solve the quadratic equation given in Eq.\ref{polaritonHAM} which yields,
 {when $Re[E_{TS}]>Re[E_{DB}]$,
 \begin{equation}
 \label{Bethe}
 W_{\pm}^{polariton}= E_{DB}-i\frac{\Gamma_{TS}+\Gamma_{DB}}{4}\pm\sqrt{-(\frac{\Gamma_{TS}-\Gamma_{DB}}{4})^2+4\alpha_{cav}^2d_{{TS,DB}}^2}
 \end{equation}
 Emission of a photon with the energy of $\hbar\omega_{cav}$ enables the coupling between the TS and the the DB resonances.}
  Upon the situation that the discriminant is vanished  the rate of the equation is reduced since $\Gamma_{TS}<<\Gamma_{DB}$. However as $\epsilon_{cav}$ is increased (and therefore also $\alpha_{cav}$) then the discriminant is a positive number  when the dipole transition is real, and consequently the upper polariton decays approximately {\color{black}  as at the exceptional point (EP) condition. The EP corresponds to a non Hermitian degeneracy (see chapter 9 in Ref.\citenum{NHQM-BOOK}).  Such a non-Hermitian degeneracy occurs when the discriminant in Eq.\ref{Bethe} vanishes, and the decay rate of the polariton is about equal to the average value of the reaction rate in the transition state and of the excited state which decays faster. }
 This result, where the decay rate is not increased when the intensity of the coupling of the activated complex  with the quantized field modes is increased, reminds of how in 1933 Bethe explained the unusual long lifetime of hydrogen in $2s$ state in strong dc field (see Eq.~25 in Ref.~\citenum{lamb1950}).  In our case the dipole transitions is complex and therefore $$ \Gamma_{min}^{polariton}= \frac{\Gamma_{TS}+\Gamma_{DB}}{4}- 2 \sqrt{\alpha^2_{cav}Im[d_{{TS,DB}}^2]}$$ where $$\alpha^2_{cav}=\frac{\Re[d^2_{{TS,DB}}]}{(\Gamma_{TS}-\Gamma_{DB})^2}.$$
 Consequently,
 \begin{equation}
     \label{GammaMIN}
  \Gamma_{min,\pm}^{polariton}= \frac{\Gamma_{TS}+\Gamma_{DB}}{4}\pm 2 \sqrt{\frac{\Re[d^2_{{TS,DB}}]Im[d_{{TS,DB}}^2]}{(\Gamma_{TS}-\Gamma_{DB})^2}}
 \end{equation}

The calculations of the effect on the cavity on on the reaction rate is carried out under the approximation that the cavity frequency $\omega_{cav}$ is tuned to couple the TS complex pole (i.e., TS-
resonance) and one of the dynamical-barrier  complex poles, so called DB-resonance. See Fig.\ref{POLES_OD2}. This approximation holds when the strength coupling parameter between the molecular system and cavity
is weak. When the complex poles are used as a basis set the polaritons (mixture of molecular states and quantized field mode states; $|E_{res}^{TS}-\frac{i}{2}\Gamma_{res}^{TS}\rangle|0_{photon}\rangle$,  $|E_{res}^{DB}-
\frac{i}{2}\Gamma_{res}^{DB}\rangle|1_{photon}\rangle$)
the following time-independent Schr\"odeinger equation is obtained,
\begin{equation*}
\begin{pmatrix}
E_{res}^{TS}-\frac{i}{2}\Gamma_{res}^{TS}-W_\pm^{polariton} &  \alpha_{cav}d_{TS,DB} \\
\alpha_{cav}d_{DB,TS}& E_{res}^{DB}-\frac{i}{2}\Gamma_{res}^{DB}+\hbar\omega_{cav}-W_\pm^{polariton} \\
\end{pmatrix}
\begin{pmatrix}
A_\pm^{TS} \\
A_\pm^{DB} \\
\end{pmatrix}
= 0
\end{equation*}
where $$W_\pm^{polariton}=E_\pm^{polariton}(\epsilon_{cav})-\frac{i}{2}\Gamma_\pm^{polariton}(\epsilon_{cav}).$$ and
$$|Pol_\pm\rangle = A_\pm^{TS}|TS\rangle|0_{photon}\rangle+A_\pm^{DB}|DB\rangle|1_{photon}\rangle $$ are the polariton state solutions.
Notice that $\alpha_{cav}(\epsilon_{cav})$ since the $\omega_{cav}$ is held fixed when $\epsilon_{cav}$ is varied. Therefore, $\alpha_{cav}(\epsilon_{cav})$ is $\omega_{cav}$ dependent.
The probabilities of the two polaritons to populate the TS is given by $\big|A_{\pm}^{TS}\big|^2$. Therefore, the effect of the cavity on the TS complex-pole is given by
\begin{equation}
\label{ENERGYpol}
    E_{polariton}(\epsilon_{cav})=\big|A_{+}^{TS}\big|^2E_+^{polariton}(\epsilon_{cav})+\big|A_{-}^{TS}\big|^2E_-^{polariton}(\epsilon_{cav})
\end{equation}
and
\begin{equation}
\label{GAMMApol}
    \Gamma_{polariton}(\epsilon_{cav})=\big|A_{+}^{TS}\big|^2\Gamma_+^{polariton}(\epsilon_{cav})+\big|A_{-}^{TS}\big|^2\Gamma_-^{polariton}(\epsilon_{cav})
\end{equation}
The minimal value of $\Gamma_{polariton}(\epsilon_{cav})$ is obtained from Eqs.\ref{GammaMIN},\ref{GAMMApol} and is given by,
\begin{equation}
\label{minGAMMApol}
    \Gamma_{min}^{polariton}(\epsilon_{cav})=\big|A_{+}^{TS}\big|^2\Gamma_{min,+}^{polariton}(\epsilon_{cav})+\big|A_{-}^{TS}\big|^2\Gamma_{min,-}^{polariton}(\epsilon_{cav})
\end{equation}
Notice that both $\Gamma_{polariton}(\epsilon_{cav})$ and $\Gamma_{min}^{polariton}(\epsilon_{cav})$ are also functions of $\omega_{cav}$ as one can see from
 Fig.~\ref{Gamma_polOD2} which presents the rate of reaction inside the cavity  as a function of the strength of the coupling of the activated complex $[AB]^\#$ with the cavity for two different values of $\omega_{cav}.$
 { \color{black} It shows how the decay rate of the $O(triplet)+D_2\to OD+D$ reaction is  \textit{increased}  inside the cavity when the distance between the two mirrors is tuned to the resonance condition}.
 The coupling between the cavity and the system, in the acceleration gauge representation is used here, where the dipole transition is given by, $d_{{TS,DB}}(\omega_{cav})=\langle E_{TS}|\partial V_{ad}/\partial X|E_{DB}\rangle$ and the coupling strength parameter (in atomic unites)  is given by $\alpha_{cav}=\epsilon_{cav}/\omega_{cav}^2$\cite{moiseyev-sindelka-2022AG}.
 For the resonance condition we use one might be able to select several different cavity frequencies. $\omega_{cav}$ which couples the TS (transition state) resonance with a dynamical barrier (DB) broad resonance provided its position is lower than the position of the TS resonance (such that the TS in the vacuum of the cavity mode have about the same energy as the DB resonance that in the first excited quantized field mode of the cavity),
 {Notice that  the minimal distances between the two       mirrors is given by, $L_Z=\pi c/\omega_{cav}\equiv\lambda_{cav}/2$.   Therefore, the minimal distance depends on $\omega_{cav}$.

As expected by the analysis given in  Eq.\ref{Bethe}, the enhancement of the reaction rate by the cavity  is \textit{reduced} as the coupling of the molecular state with the external field is increased.
 This is a special case of the generality of Bethe's proof that there are situations where the increasing of the strength of the coupling of the system with external field not necessarily increase the rate of the reaction.
  See Fig.\ref{Gamma_polOD2} that shows the Bethe effect in the calculations of the effect of the cavity on the decay rate of the chemical reaction. That is, the increasing of the coupling strength parameter does not result necessarily in increasing the rate. When the difference between the resonance widths is sufficiently large (i.e., the TS is coupled with a broad DB resonance) the increasing of the strength of the coupling between the cavity and the molecules results in the suppression of the rate. As one can see from the results presented in Fig.\ref{Gamma_polOD2} the \textit{maximal enhancement} of the rate of the reaction is obtained for a relative weak coupling between the cavity and the molecules.\\

{\subsection{Example - The conditions that should be satisfied in order to enhance the rate of reaction in a cavity:  $O+D_2\to [ODD]^\# \to OD+D$ as a feasible example.}}

Using the adiabatic resonance eigenvalues and eigenfunctions for the $O+D_2$ reaction we have calculated and introduced in the previous section we evaluate the effect of the cavity on the reaction rate using Eq.\ref{minGAMMApol}.

   \begin{figure}[h!]
     \centering
     \hspace*{+0.9cm} 
     \includegraphics[angle=000,scale=0.5]{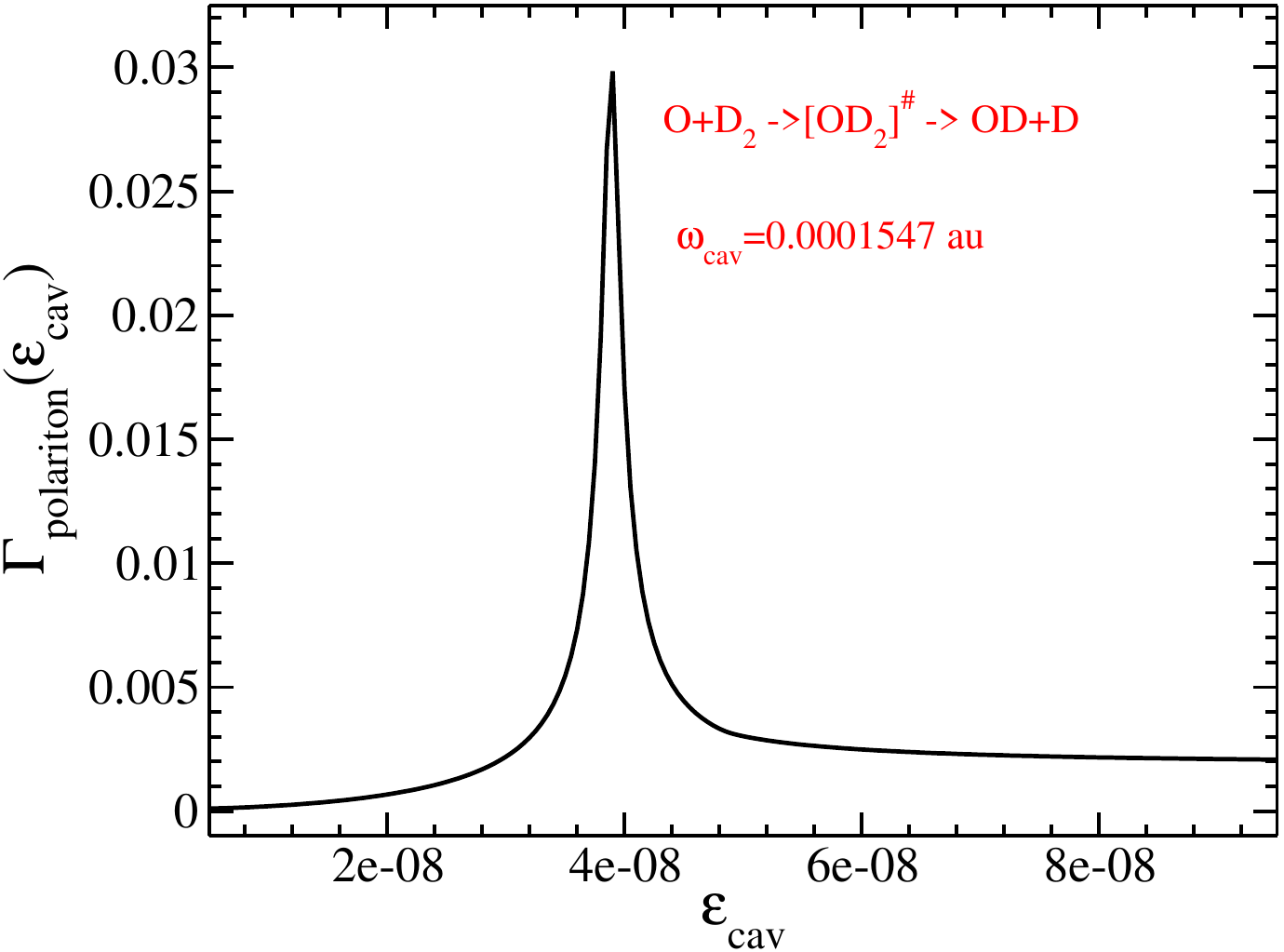}
      \caption{ {The effect of the cavity on the reaction  rate, $\Gamma_{polariton}$,  of  the TS polaritions of $O+D_2 \xrightarrow[]{\Gamma_{polariton}(\epsilon_{cav})} OD+D $ reaction. Reaction rate in Hartree and cavity-coupling parameter, in atomic units.  The cavity be tuned into resonance condition between the states marked in blue in Fig.\ref{POLES_OD2} by adjusting the distance between the two Fabry-Perot mirrors (see Eq.~\ref{epsilon_VOL_cav}). Rate in Hartree and coupling in atomic units. {\color{black} Notice that outside of the cavity (i.e., $\epsilon_{cav}=0$)
      the reaction rate is very small compared to the setup when the cavity is present.
 {\textit{ In other words, in this case the cavity effect on increasing the reaction rate is enormous even when the strength of the coupling between a single molecule and the cavity is weak}.}}}
      See Eq.~\ref{epsilon_VOL_cav}, which links
      $\epsilon_{cav}$ to the distance between the two mirrors of the Fabry-Perot dark cavity.}
    \label{Gamma_polOD2}
    \end{figure}
\newpage

\section{The influence of a dark cavity on additional chemical reactions }
\label{arhcl}
An example for the first case case is scattering of  hydrogen atom from the Van der Waals complex ArCl\cite{garcia-Gerber-Valantiny-1991}. The poles which are presented in Fig.2 of Ref.\citenum{Eddy-NM-ArHCL-Mol-Phys-1998} are resonances which are associated with the temporary trapping
of a hydrogen atom in between Ar and Cl. Only the first row of resonances in this figure
have a classical analogue and are associated with the
classical unstable periodic orbits. All other resonances
are quantum diffraction resonances which do not have a
classical analogue. These are Feshbach resonances of the two-dimensional PES where one coordinate is the reaction coordinate and the second one is perpendicular to it.
These Feshbach resonances were found to be in excellent agreement with the complex poles of the adiabatic one-dimensional potential as described here in Eq.\ref{ADIBATIC-PES}. The narrowest resonance is located at the top of the DB potential and describe the TS. Beside
the physical insight we get by the use of the adiabatic
approach we also gain in the drastic reduction of the
computational effort which is required for calculating
the poles. Here we have shown that the non-resonance poles of
1D Eckart-type potential barrier (associated with complex scaled eigenfunctions which are localized in the
forbidden classical region in 2D phase space) are the
resonance poles of the 2D problem and are associated
with complex-scaled eigenfunctions which are localized
in the classically accessible region in the four-dimensional phase space. These type of resonances are marked here as DB resonances.  The idea is that the Fabry-Perot cavity couples the TS resonance with one of the DB resonances which have a broader width then the width of the TS and thereby the rate of the reaction $Ar+HCl\to H+Ar+Cl$ is enhanced.

\section{Possible Feasible Experiments for Enhancing the Rate of Gas-Phase Reactions in a Dark Cavity}
\label{cons}

The conditions for enhancing the rate of gas-phase reactions by a cavity are as follows:

(I) A necessary (but not sufficient) condition is that the reaction will be controlled by the Feshbach resonances of the activated complex within multidimensional space. This condition manifests as the emergence of a dynamical barrier in the 1D potential under the adiabatic approximation. The adiabatic approximation is valid when the reaction along the reaction coordinate is slower than the motions along coordinates perpendicular to the reaction coordinate.

There are two scenarios in which this condition is satisfied:

(a) When the sum of the bound vibrational frequencies of the activated complex (transition state), which are perpendicular to the reaction coordinate, exceeds both the sum of the normal modes of the reactants and that of the products. In addition, the barrier along the reaction coordinates will either be negligible or sufficiently broad and large to facilitate an adiabatic transition of the vibrational modes from the reactants through the activated complex to the products.

(b) When the sum of the bound vibrational frequencies of the activated complex (transition state) is slightly lower than the sum of the normal modes of the reactants and that of the products, forming an asymmetric potential well with dynamical barrier (DB) resonances. In this scenario, the transition state (TS) is associated with the narrowest resonance, located near the top of the dynamical potential barrier, and the  DB resonance that its position is lower than the position of the TS resonance has a larger width (lower lifetime).

The distance between the two mirrors is verified from a large distance where the interaction of the molecule with the cavity is practically zero, to an optimal distance where the standing quantized field mode has the same wavelength as the wavelength that couples the transition state (TS) complex pole and the dissociation barrier (DB) complex pole (which is inactive when the two mirrors are far apart).

\textit{Condition (1a) is satisfied}, for example, in the scattering experiments of hydrogen atoms from the Van der Waals molecules ArCl. The products are Ar+H+Cl \cite{Eddy-NM-ArHCL-Mol-Phys-1998}.
In this case the potential barrier along the reaction coordinate is negligible and the only significant potential barrier is the dynamical potential barrier (DB). \textit{As we have shown here this reaction will be enhanced  in dark cavity}.

\textit{ Condition (1b)} is satisfied, for example, in the reaction $OH+ HO_2 \leftrightarrow O_2+H_2O $  (triplet reaction). See Table in Ref.\citenum{2019-2ndEXAMPLE} where the vibrational energies perpendicular to the reaction coordinate  at the two almost degenerate transition points (TS1 and TS2) are given and also the normal vibrational modes of $OH+HO_2$ and of $O_2+H_2O$.  In this example, two nearly degenerate adiabatic potential energy surfaces are illustrated, as shown  in Fig.~4 of Ref.\citenum{dynamical-barriers-2023}. In this work  we have shown that for asymmetric reaction of $O+D_2$} the rate of the reaction is {\color{black}\textit{significantly increased
 by the dark cavity}.} \\

 \section{Concluding remarks}
{In this work we investigate the conditions for enhancing the rates of reactions within a dark cavity.
We use the  adiabatic approximation by describing the multi-dimensional PES  using a reaction path Hamiltonian, where $X$ is the reaction coordinate and $\omega_j(X)$ are the frequencies of the normal vibrational modes which are perpendicular to the direction the reaction  coordinate $X$.
 The one-dimensional potential is given by,
 $$V_{r.c}(X)+\hbar\sum_{j=1}^{3N-7} \omega_j(X)(n_j+\frac{1}{2}),$$ where $n_j=0$ (ground vibrational states).
 $V_{r.c}(X)$ is  the static potential barrier for the reaction coordinate trajectory, i.e., it is solutions of the many-body electronic Hamiltonian.}
 This approach has been previously used  to study the collisions of hydrogen atom with $ArCl$ Van der Waals molecule\cite{Eddy-NM-ArHCL-Mol-Phys-1998} and for several collision  reactions, such as in Ref.\citenum{2019-2ndEXAMPLE}.
 {Moreover, NHQM is used to describe the resonance states (complex poles of the scattering matrix) that are used as a basis set within the adiabatic approach.)}

 {Conditions for enhancing reaction rates are derived and presented in Section~\ref{cons}.}
 Using the first  condition  we show that for the collision of hydrogen with the van der-Walls molecule the rate of the reaction is enhanced significantly  inside a tuned dark cavity.
  Notice that there are no photons in the cavity, in addition, the reactants are in their ground electronic and vibrational states, i.e., it is indeed a dark cavity.
 {These conditions have been  applied  also to the collision reaction of oxygen (in a triplet state) with deuterium diatomic molecule.
 We demonstrate a dramatic enhancement of the reaction rate in a dark cavity, where the distance between the two mirrors in the Fabry Perot cavity are tuned.
 That is, the quantized field mode state energy, $\hbar\omega_{cav}$, equals the energy difference between two coupled resonance states.
  Finally, based on our analysis for the \textit{symmetric}  hydrogen exchanged reaction in methane the rate of the reaction cannot enhanced by a dark cavity. {\color{black}We have not presented details of this analysis, since the adiabatic 1D effective potential and the 1D potential obtained from electronic structure calculations along the 1D curvilinear reaction coordinate are almost identical, and, therefore, no effect of the cavity on the reaction rate can be observed.}
  Hence here we provide general guidelines for enhancing the rates of reactions inside a dark cavity.

\acknowledgments{ Dr. Arie Landau from the Helen Diller Center for Quantum Mechanics at the Technion is acknowledged for carefully reading several versions of this manuscript and for his valuable comments and suggestions to simplify the representation of the conditions that should help identify which reaction rates can be enhanced in a dark cavity.  {\color{blue} Dr. Suheir Assady from RAMBAM is acknowledged for her hospitality in her department during the major portion of this research. Besides providing me with privacy, she also gave me the peace of mind necessary to conduct my scientific work by applying to biology what is known in physics and chemistry as the von Neumann and Wigner Non-Crossing Rule. This rule suggests that when several unexpected phenomena are observed simultaneously, they are most likely correlated, and their correlation typically requires the optimization of at least two parameters. } }

\bibliography{References}

 \end{document}